\documentclass[preprint]{aastex}
\usepackage{epsfig,emulateapj5}

\newcommand{\km}{${\rm km\,s}^{-1}$}

\shorttitle{Observations of the Magellanic Bridge gas}
\shortauthors{Lehner et al.}

\begin{document}

\title{Observations of the interstellar medium in the Magellanic Bridge\altaffilmark{1}}

\author{N. Lehner\altaffilmark{2,4}, K.\,R.~Sembach\altaffilmark{3}, P.\,L.~Dufton\altaffilmark{2},
W.\,R.\,J. Rolleston\altaffilmark{2} and F.\,P.~Keenan\altaffilmark{2}}

\altaffiltext{1}{Based on observations with the NASA/ESA Hubble Space Telescope,
obtained at the Space Telescope Science Institute, which is operated by the
Association of Universities for Research in Astronomy, Inc. under NASA
contract No. NAS5-26555.}

\altaffiltext{2}{Department of Pure and Applied Physics, The Queen's University
of Belfast, Belfast, BT7 1NN, Northern Ireland}

\altaffiltext{3}{The Johns Hopkins University, Department of Physics and Astronomy,
Bloomberg Center, 3400 N. Charles Street, Baltimore, MD 21218, USA}

\altaffiltext{4}{Present address: The Johns Hopkins University, Baltimore, MD 21218, USA}

\begin{abstract}
We present  ultraviolet and optical spectra of DI\,1388,  
a young star in the Magellanic Bridge, a region
of gas between the Small and Large Magellanic Clouds.
The data have signal-to-noise ratios of 20--45 and a spectral
resolution of 6.5 \km.
Interstellar absorption by the Magellanic Bridge  at 
$v_{\rm LSR} \approx 200$ \km is visible in the lines of  \ion{C}{1},
\ion{C}{2}, \ion{C}{2*}, \ion{C}{4}, \ion{N}{1}, \ion{O}{1}, \ion{Al}{2},
\ion{Si}{2}, \ion{Si}{3}, \ion{Si}{4}, \ion{S}{2}, \ion{Ca}{2},
\ion{Fe}{2}, and \ion{Ni}{2}. The relative gas-phase abundances of \ion{C}{2}, 
\ion{N}{1}, \ion{O}{1}, \ion{Al}{2}, \ion{Si}{2}, \ion{Fe}{2}, and \ion{Ni}{2} 
with respect to \ion{S}{2} are similar to those found in Galactic halo 
clouds, despite a 
significantly lower metallicity in the Magellanic Bridge. 
The higher ionization species in the cloud have a column density ratio
$N({\rm C^{+3}})/N({\rm Si^{+3}}) \sim 1.9$, similar to that inferred for 
collisionally ionized Galactic cloud interfaces at temperatures $\sim 10^5$ K.
We identify sub-structure in the 
stronger interstellar lines, 
with a broad component (FWHM $\sim 20$ \km) at  $\sim 179$ 
\km and a sharp component (FWHM $\sim 11$ \km) at 198 \km. The 
abundance analysis for these clouds indicates that the feature at 198 
\km consists of a low electron density, mainly neutral gas that may
be associated with an interface responsible for the highly ionized 
gas.  The 179 \km cloud consists of  
warmer, lower density gas that is partially ionized. 
\end{abstract}


\section{Introduction}
The Large (LMC) and Small (SMC) Magellanic Clouds are two small irregular
galaxies in orbit around the Galaxy, with the LMC being approximately 3 to
10 times more massive than the SMC \citep{dopita90}. Distance
estimates for the LMC and SMC are $\sim 49$ and $\sim 60$ kpc, respectively
(see e.g., van den Bergh 1999 and references therein).
Evidence for interactions between
the Milky Way and the Magellanic Clouds is provided by several 
high velocity  gas complexes connected to the Clouds: the Magellanic
Bridge (MB), a $10\degr$ region of \ion{H}{1} linking the body 
of the SMC to an extended arm of the LMC; the Magellanic
Stream, a $10\degr \times 100\degr$ \ion{H}{1} filament that trails the 
Clouds; and the Leading Arm, a diffuse \ion{H}{1} region that leads 
the Galaxy and the Clouds
\citep{putman}. Using data from the \ion{H}{1} Parkes All-Sky Survey (HIPASS), 
\citet{putman} showed that the ``boundary'' between the SMC and the MB occurs
at $l = 295 \degr$, $b = -41.5\degr$, where the \ion{H}{1} column density is 
$\sim10^{21}$ cm$^{-2}$. The \ion{H}{1} column density
slowly decreases toward the LMC to $\sim10^{20}$ cm$^{-2}$ at the LMC 
boundary near $l = 287 \degr$, $b = -35.5 \degr$.  The LSR velocity 
of the MB gas observed in 21\,cm emission ranges from $\sim100$ to 350 \km 
\citep{mcgee}.

Various formation mechanisms for the Bridge, Stream, and Leading Arm 
have been suggested over the years, and it is now generally agreed that 
the MB formed via a tidal encounter between the two Clouds. \citet{gardiner}
have produced models that can reproduce simultaneously both the MB and 
the Stream observations.  They  
find that the Bridge was most likely pulled from
the wing of the SMC 200 Myr ago during a close encounter between the 
two Clouds, whereas the Stream was created by a SMC-LMC-Galaxy close 
encounter $\sim1.5$ Gyr ago. 

Searches for stars in the Stream have produced largely negative 
results (see Irwin et~al. 1990 and references therein). However, photographic 
surveys employing automatic scans of Schmidt plates have now firmly 
established the existence of blue stellar objects within the MB 
\citep{irwin90}.
Subsequent CCD photometry \citep{demers} and spectroscopy \citep{rol99}
have shown that the MB contains massive, young ($<20$ Myr) stars located
between the LMC and SMC. 
These stars have a metal abundance $\sim 1.1 $ dex lower than their Population
I Galactic analogues and $\sim 0.5 $ dex lower than the SMC
(Hambly et al. 1994; Rolleston et al. 1999). Their evolutionary 
lifetimes indicate that star formation is still occurring in the MB, but 
probably via different mechanisms than those in the Galaxy, as the MB has a 
very low \ion{H}{1} column density (but note the recent detection of cool 
\ion{H}{1} clouds in the MB by \citet{kob}, indicating that some regions 
have relatively high densities).

The combination of high spectral resolution
and high sensitivity available with
space-based ultraviolet telescopes makes it feasible to 
investigate the chemical composition and the physical conditions
in various interstellar environments in the Galaxy and Magellanic Clouds 
(see Savage and Sembach 1996 for a review).  Here, we report the results 
of a programme to investigate the chemical composition and abundance pattern 
of the MB gas with the Hubble Space Telescope {\em (HST)}
and the Space Telescope Imaging Spectrograph (STIS).  For this investigation,
we have chosen a sight line for which \ion{Ca}{2} absorption was detected 
previously by \citet{hambly}.

\section{Observations and data reductions}
\subsection{DI\,1388}
Previously, we obtained 
high resolution optical spectra for a number of B-type stars in the 
Magellanic Bridge using the 3.9\,m Anglo-Australian Telescope (AAT), during 
observing runs in 1992 November and 1995 December. One of the stars 
observed by \citet{hambly}, DI\,1388, is situated approximately mid-way 
between the SMC and LMC ($l = 291.2\degr$, $b = -41.2\degr$). 
Interstellar \ion{Ca}{2} K absorption at a LSR velocity of
$200$ \km has been detected toward DI\,1388 \citep{rol99}.
This  velocity is consistent with that of
the Bridge material seen in \ion{H}{1} 21 cm emission \citep{mcgee,putman}.
DI\,1388 has  relatively weak optical
stellar absorption lines and a high projected rotational velocity. 
DI\,1388 has the weakest  MB ISM \ion{Ca}{2} absorption in
the \citet{rol99} sample ($W_\lambda \sim 42$ m\AA).

Recent work
(Ryans, priv. comm.) classifies DI\,1388 as a main-sequence, very early 
B-type or late O-type star.  In Table~\ref{t0} we summarize its atmospheric 
and other observational parameters. 
\subsection{{\em HST}\, STIS spectra}\label{hstdr}
{\em HST}\, STIS spectra of DI\,1388  were obtained on
1997 November 21.
The data are catalogued in the HST archive under the 
identification O4A801010. A total exposure time of 26,160 seconds
was obtained with the far-UV \'echelle spectrograph 
(FUV, $1150-1730$ \AA)
using the FUV-MAMA detector in the ACCUM mode. The E140M grating was employed 
centered at 1425 \AA, with a $0\farcs2\times 0\farcs06$ slit in order to obtain
the highest resolution available with this grating. This configuration 
resulted in a two-pixel (FWHM) resolution of $\sim 6.5$ \km. Table~\ref{t3}
summarizes the ions observed toward  DI\,1388 and the respective S/N levels in
the nearby stellar continuum.

Data were reduced within the {\sc noao iraf} package \citep{morris}
using the {\sc stsdas} packages. Standard  calibration and extraction
procedures were employed using the {\sc calstis} routine.  The spectra 
from the four different exposures  
were co-added. Spectral regions containing interstellar lines
were normalized by fitting low-order polynomials
($\leq 4$, varying with the complexity of the continuum shape) to 
continuum regions. No special processing to adjust the background 
levels was necessary (e.g., Howk and Sembach 2000). The cores of
the strongly saturated interstellar lines of \ion{C}{2} $\lambda1334$,
\ion{O}{1} $\lambda 1302$, and \ion{Si}{2} $\lambda1304$ are within 1--2\% 
of the zero-flux 
level, indicating that the uncertainties introduced by scattered light 
are small for the weaker absorption features analyzed in this study
(see Figure~\ref{fig2}).
At wavelengths shortward of Ly$\alpha$ spurious structures might be present
near some lines (e.g., \ion{N}{1}, \ion{Si}{2}), but the S/N is lower and it 
is possible that some of these features are simply random noise fluctuations.
\subsection{AAT observations}
Subsequent high resolution optical observations were undertaken on 1999 
October with the AAT to obtain both a high quality spectrum of the 
interstellar \ion{Ca}{2} K line and to detect stellar features.
The coud\'e University College London Echelle Spectrograph (UCLES) was employed
with a Tektronix CCD detector, and the dispersing element was a 31.6 lines/mm
grating. Bias and quartz lamp flatfield exposures were taken at the 
start and end of each night. Additionally, ThAr exposures were
interleaved with the observations to allow wavelength calibration
and  to determine the spectral resolution of the data. This instrument
configuration resulted in a FWHM velocity resolution of $\sim 7.1$ \km and a
signal-to-noise ratio of 45 in the continuum.

The optical data were reduced within the {\sc noao iraf}
package. Standard  procedures were
followed in all cases, with images being debiased, flatfield corrected, trimmed
and cleaned of cosmic rays. The spectra were traced and sky subtracted.
Wavelength  calibration was performed by identifying the locations of 
emission lines in the relevant arc spectra.
The one dimensional spectra were normalized by fitting a low-order polynomial 
to continuum regions. Cross-correlation techniques were used to align the
individual spectra. The spectra were then weighted by the continuum 
signal-to-noise levels and co-added.

\section{Analysis}
\subsection{Spectra and equivalent widths}\label{ew}
We transformed the spectra to the dynamical LSR \citep{mihalas} 
using corrections obtained from the {\sc starlink rv} package
\citep{wallace}. 
We then modeled the interstellar lines using the optimized Gaussian fitting 
routines in the {\sc starlink dipso} package \citep{howarth} to obtain radial 
velocities  and equivalent widths. The correction for the HST spectra in the 
direction of DI\,1388  is $v_{\rm LSR} - v_{\rm helio} = 10.7$ \km. The MB 
\ion{H}{1} emission has LSR velocities between 100 and 350 \km. Thus, we 
assumed that 
absorption at LSR velocities lower than 100 \km is from the 
Milky Way disk and halo, while that at 
$v_{\rm LSR} = 200 $ \km was assumed to arise in the Magellanic Bridge 
(see Figure~\ref{fig2}). 
In the {\em HST}\, STIS DI\,1388 spectra, a high-velocity cloud (HVC) 
absorption is present at $v_{\rm LSR} = 79.7 \pm 3.4$ \km (see Lehner 
et al. 2000).  In Figure~\ref{fig2}, we note the detection of very weak 
features in the spectra of \ion{C}{2}
$\lambda1334$ and \ion{Si}{2} $\lambda\lambda$1193 and 1260 at LSR 
velocities of 113 and 130 \km. The velocities of these two weak HVCs
suggest that they are associated with the MB \citep{lehner1}.

To check the consistency between absorption line velocities in the optical and 
ultraviolet absorption spectra, we compared the low velocity portions of the
\ion{Ni}{2} $\lambda$1370 and \ion{Ca}{2} K lines.  These two lines have 
roughly the same strength and component structure.
The difference in the LSR velocities of the two lines is less than 3 \km,
which is less than the instrumental resolution of either dataset.  
The absorption profile velocities also agree well with those of 
\ion{H}{1} emission observations obtained with the ATCA \citep{lehnerb} and 
Parkes (Putman, priv. comm.) telescopes.
As a final check, we compared the stellar radial velocities in both the 
optical and ultraviolet spectra of DI\,1388 and found them to be consistent 
within their uncertainties.

Table~\ref{t3}  presents the equivalent width results 
for the MB absorption.  
The S/N ratios are sufficient to detect two components in many of the 
stronger lines arising in the MB; this structure manifests itself as the 
combination of a broad component at 179 \km and a 
narrow component at 198 \km (see Figures~\ref{fig2} and \ref{fig3} 
and discussion in Section~\ref{cfm}). 
Owing to the large stellar projected 
rotational velocity, contamination by stellar absorption is generally 
unimportant for the MB absorption, except for the higher ion lines of 
\ion{C}{4}, \ion{Si}{3} and  \ion{Si}{4}, as illustrated in Figure~\ref{fig2}.

Errors on the  equivalent widths in Table~\ref{t3}
are $1\sigma$ estimates, including statistical noise fluctuations in the 
lines and systematic errors arising from continuum placement 
uncertainties. The latter were obtained by 
changing the continuum by an amount equal to $\sim 0.2 - 0.4 $ times the 
RMS noise value, where the scale factor 0.2 corresponds to a flat continuum 
and 0.4 to a continuum with large curvature \citep{sembach92}. 
The $3\sigma$ upper limits for the equivalent widths 
are defined as $W_{\rm min} = 3 \sigma\, 2\delta\lambda$, where 
$\sigma$ is the inverse of continuum S/N ratio, $\delta \lambda$ 
is the spectral resolution in m\AA, and $2 \delta \lambda$ 
reflects approximately the FHWM of the absorption lines (see Table~\ref{t4}).
Other errors such as  background and scattered light uncertainties
are not included but are expected to be small since the zero-level 
is relatively well determined (see Section~\ref{hstdr}).

\subsection{Apparent column density profiles}
We used the apparent optical depth method to derive column 
densities and to check for unresolved saturated structures within the observed 
profiles \citep{savage}.  
We converted the normalized absorption profiles into apparent
optical depths
per unit velocity, $\tau_a(v) = \ln[1/I_{\rm obs}(v)]$, where $I_{\rm obs}$ 
is the normalized observed intensity. These values of  $\tau_a(v)$ are related
to the apparent column densities per unit velocity, $N_a(v)$ 
(cm$^{-2}$ (\km)$^{-1}$) through the relation 
$ N_a(v) = 3.768 \times 10^{14} \tau_a(v)/[f \lambda(\rm\AA)]$.
A direct integration of the apparent column density profiles over the 
velocity range yields the total column densities of the lines, provided 
there are no unresolved saturated structures present
\citep{savage}. We adopted wavelengths and oscillator strengths from 
the \citet{morton} atomic data compilation unless otherwise 
indicated in Table~\ref{t3}.

The last column of Table~\ref{t3} gives the resulting column 
densities\footnote{In this table 
\ion{N}{1} $\lambda$1199 and \ion{Fe}{2} $\lambda$1260 column densities were obtained by component 
fitting to deblend the lines from lower velocity components and other species.} 
for the different ions in the MB gas 
observed toward DI\,1388. Errors  on the column densities
are $1 \sigma$ estimates (see the equivalent width uncertainties in 
Section~\ref{ew}).
Upper limits ($3\sigma$) are
obtained from the corresponding equivalent width limit and the assumption 
of a linear curve of growth. Lower limits indicate that the line 
contains some unresolved, saturated absorption that cannot be reliably 
estimated with the existing data. In the absence of unresolved saturated 
structure, two or more lines of a given species with different values of 
$f\lambda$ will have the same distribution of $N_a(v)$. Differences in 
$N_a(v)$ suggest that some saturation is present for the stronger line(s).  
Figure~\ref{fig4} clearly demonstrates this for the MB \ion{Si}{2} lines;
no unresolved saturated absorption is observed for 
\ion{Si}{2} $\lambda$$\lambda$1304 and 1526, while $\lambda$1190 is slightly 
saturated and $\lambda$1193 is strongly saturated. The other singly ionized 
species follow a similar curve of growth (similar $b$-values -- see 
Section~\ref{cfm}), and therefore the weaker singly ionized lines are not saturated.
\ion{C}{2} $\lambda$1334
and \ion{Si}{3} $\lambda$1206 are stronger lines that may contain 
some unresolved saturated structures at the MB velocity. 
The \ion{N}{1} lines
have lower effective $b$-values that place them on the flat part of the curve of growth.
This is confirmed in Figure~\ref{fig4}, where comparison of the apparent 
column density profiles for the different lines
shows that some unresolved saturated structure is present (as expected for 
narrower features).
\ion{O}{1} $\lambda$1302, which probably follows a curve of growth similar 
to that of the singly ionized species, is strong and probably contains
some unresolved saturated structure.

\subsection{Component-fitting measurements}\label{cfm}
For the relatively strong lines of \ion{C}{2}, \ion{N}{1}, \ion{O}{1}, 
\ion{Al}{2}, \ion{Si}{2}, \ion{S}{2}, \ion{Fe}{2},
and the \ion{Ca}{2} line, we also used a Voigt profile 
fitting method \citep{welty91} to measure the column densities and the 
$b$-values. We modeled these features first with
the optimized Gaussian fitting routines ({\sc elfinp}) in the  {\sc
starlink} package  {\sc dipso}  \citep{howarth} 
to obtain radial velocities, equivalent widths, and the full
widths at half maximum intensity (FWHM$= 2 \sqrt{\ln2}\, b$, where 
$b^2=2kT/(A m) + v^2_{\rm turb}$,
and $A$ and $m$ are the ion mass in atomic mass units and the hydrogen mass, 
respectively).
We then used two of the three parameters from the fit ($v_{\rm LSR}$, $b$) as 
initial estimates in the Voigt profile fitting programme. The results of this 
process are 
summarized in Table~\ref{t4}, and some spectra with their respective fits are 
presented in Figure~\ref{fig3}.
The errors in this table are $1\sigma$ estimates, including statistical noise 
fluctuations in the lines and systematic errors due to uncertainties in the 
continuum placement (see previous section). We note that the component fit results are 
consistent with the apparent optical
depth column densities.

The \ion{C}{2}, 
\ion{O}{1}, \ion{Al}{2}, \ion{Si}{2}, \ion{S}{2},
and \ion{Fe}{2} features have a two-component structure,
with a broad feature centered at $\sim 179$ \km and sharp component 
at $\sim 198$ \km. 
The fit for the strong \ion{C}{2} line is uncertain because of
the strength of the 
line.  The fit for the \ion{Al}{2} line is also uncertain since the 
line falls near the edge of an \'echelle order and is not completely
covered in these data. The $b$-values for \ion{S}{2} appear 
to be smaller than for the  other absorption lines. However, for $b$-values
between 7 and 10 \km (typical values for the other lines), the derived \ion{S}{2} column 
densities do not change by more than the $1\sigma$ error indicated in 
Table~\ref{t4}.  The residuals in the fit for \ion{O}{1}
indicate that the sharp component contains either some unresolved saturated 
structures or additional weak components.  The 
column density of the broader \ion{O}{1} component is also somewhat 
uncertain as it lies on the flat part of the curve of growth.
The \ion{N}{1} lines exhibit a well-defined sharp component at $\sim 198$ \km, 
but the broad component is not detected in the $\lambda$1200 line and is very 
weak in the $\lambda$1201 line.
The \ion{N}{1} $\lambda$1200 non-detection could be due to the continuum 
uncertainties on the blue side of this line (see Figure~\ref{fig2}).  
For the weak \ion{C}{2}* and \ion{Ni}{2} lines, 
the S/N ratios of the data allow identification of only the sharp component. 
Finer structure seen in higher resolution spectra of 
Galactic HVCs \citep{ryans,lehner99} suggests that the true component 
structure in the MB gas toward DI\,1388 may be more complex than is revealed
by our data.

\section{Overview of species observed in the Magellanic Bridge}
The wide range of ionization states observed for the MB shows that 
the ionization structure of these clouds is complex, as the clouds exhibit
species ranging from neutral gas tracers 
(e.g., \ion{O}{1} and \ion{N}{1}) to triply 
ionized gas tracers (e.g., \ion{Si}{4} and \ion{C}{4}). Before 
interpreting the derived column densities,
we present a short description of the observed species
and the roles that they might play within the neutral and ionized environments.

{\underline {\bf - \ion{C}{1}, \ion{C}{2}, \ion{C}{2*}, \ion{C}{4}:}}
The MB component contains a strong, saturated \ion{C}{2} $\lambda$1334 
absorption feature and a moderately strong \ion{C}{2*}  $\lambda$1335 line.  
Individual \ion{C}{1} lines are not detected at significant levels. We 
weighted the \ion{C}{1} lines by their $gf$ values and S/N levels and co-added
the lines to produce an average \ion{C}{1} line.  This resulted in a
tentative detection at 198.1 \km (Figure~\ref{fig9}); the derived
column density is $\approx12.26$ dex.
The interstellar \ion{C}{4} lines are blended with a 
broad, deep stellar line centered at $\sim 150$ \km.  We approximated
the stellar \ion{C}{4} absorption with a Gaussian profile to produce the 
interstellar column density listed in Table~2.  The effect of this deblending is 
shown in Figure~2.
\ion{C}{2} is the dominant ionization stage in both the warm neutral medium 
(WNM) and warm ionized medium (WIM)
\citep{sembach95,sembach00}, while \ion{C}{4} is primarily 
produced by collisional ionization at $T_e \sim 10^5$ K \citep{suth}. 
The combination of \ion{C}{1}, \ion{C}{2} and \ion{C}{4}
absorption in the Bridge suggests that there may be interfaces between hot 
and cool regions.

{\underline {\bf -  \ion{N}{1} and \ion{O}{1}:}}
\ion{N}{1} and \ion{O}{1} are excellent tracers of neutral gas as
their ionization potentials and charge exchange reactions with hydrogen 
ensure that they are primarily found in \ion{H}{1} regions. 
Magellanic Bridge \ion{N}{1} is observed at 1199, 1200 and 1201 \AA, but the
1199 \AA\ line is blended with the saturated local absorption in the
\ion{N}{1} $\lambda$1200 line. The mean column density of \ion{N}{1} in the MB
is greater than 14.0 dex. \ion{O}{1} $\lambda1302$ is also detected in the MB and is very 
strong.

{\underline {\bf - \ion{Si}{2}, \ion{Si}{3} and \ion{Si}{4}:}}
These ions are observed at the velocities of  the MB.
\ion{Si}{2}
is the dominant ion in both the WNM and WIM \citep{sembach00}.
For \ion{Si}{2} $\lambda$$\lambda$1304
and 1526, we have adopted the oscillator strengths of 
\citet{dufton83,dufton92}, as recent studies indicate that they are more 
accurate than other estimates \citep{spitzer}, and indeed the derived
column densities for the different \ion{Si}{2} lines agree well.
Strong \ion{Si}{3} $\lambda$1206 absorption present at MB
velocities is partially blended with a broad stellar feature.  Removing  
the deepest possible stellar line centered at 150 \km, we find a column 
density of $> 13.45$ dex for the MB \ion{Si}{3} absorption 
(see Table~\ref{t3}). The \ion{Si}{3} line shape differs substantially
from that of the singly ionized species in that the the 198 \km component is 
much broader ($b \sim 16$ \km).  The presence of \ion{Si}{3} suggests
that the gas could be partially ionized.  \ion{Si}{4} is also present in the 
MB, indicating that some hotter ($T \sim 10^5$ K) gas may also exist.
We estimated the \ion{Si}{4} line strengths by deblending the stellar
absorption using the same strategy employed for \ion{C}{4}.
The detection of \ion{Si}{3}, \ion{Si}{4}, and \ion{C}{4} at velocities 
similar to the neutral and low-ionization species indicates that the hot 
component is not circumstellar.

{\underline {\bf - \ion{S}{2} and \ion{S}{3}:}}
Sulfur is found mainly in the gas phase \citep{savage96} and can be used as 
a reference to
study the depletions of other elements 
in the gas. The \ion{S}{2} $\lambda$$\lambda$1250 and 1253 lines are weak but 
definitely detected, while the feature at 1259 \AA\ is blended with the strong 
saturated local \ion{Si}{2} $\lambda$1260 absorption. \ion{S}{3} is not 
detected.

{\underline {\bf - \ion{Al}{2}, \ion{Ca}{2}, \ion{Fe}{2}, \ion{Ni}{2}:}}
These species are grouped together as they have the general property of being
readily depleted on to dust grains \citep{savage96}. With the exception 
of \ion{Ca}{2}, 
these species are dominant in the neutral and ionized gas 
(though \ion{Fe}{3} can be
dominant in some ionized gas -- see e.g. Sembach et~al. 2000). Only part of 
the \ion{Al}{2} $\lambda1670$ MB feature is
present in the {\em HST}\, STIS spectrum, so it is difficult to measure
the line strength accurately.
\ion{Fe}{2} $\lambda$$\lambda$1260 and 1608 are detected, while $\lambda$1611 
is not detected. For \ion{Fe}{2} $\lambda$$\lambda$1608 and 1611, we 
have adopted the oscillator strengths of \citet{mullman} and \citet{cardelli}, 
respectively. Only the measured 
column density of $\lambda$1608 is reliable, as \ion{Fe}{2} $\lambda$1260 
is strongly blended with
the saturated MB component of \ion{Si}{2} $\lambda$1260. 
The \ion{Ni}{2} absorption lines are very weak. For \ion{Ni}{2}, recent 
studies \citep{fedchak,zsargo}
suggest that the oscillator strengths are lower by a factor $0.53$ 
compared to the previous values;
we have therefore adopted this scaling factor. To
improve the reliability of \ion{Ni}{2} column density, the three spectra 
were co-added using a method  similar to that employed for \ion{C}{1}. The result
is shown in Figure~\ref{fig9}, where the centroid of the line (198.4 \km) 
is in good agreement with the centroids of the stronger lines. The $b$-value 
is about 5.7 \km, a bit smaller than for the other singly ionized lines.
The derived \ion{Ni}{2} column density is $12.52 \pm 0.20$ dex.

\section{Physical conditions within the Magellanic Bridge gas}
\subsection{Depletions}
To determine 
the depletion pattern of the MB, we compared relative heavy element column 
densities. 
Since \ion{S}{2} is only modestly depleted in the Galactic ISM, we
used it as the reference ion. 
The logarithmic normalized gas-phase abundance is defined using the following 
notation:
\begin{equation}\label{dg}
[X/{\rm S}] = \log\left(\frac{X^i}{{\rm S^+}}\right) - \log\left(\frac{X}{{\rm S}}\right)_{\rm c} \,,
\end{equation}
where $X^i$ is the ion under consideration, and $(X/{\rm S})_{\rm c}$ is the ratio for cosmic 
abundances and $X/{\rm S}$ is used for $N(X)/N({\rm S})$. Equation~\ref{dg} 
assumes that the ion $X^i$ is the dominant form of element $X$. 
When it is not, we will consider explanations other than depletion on to 
dust grains (such as 
ionization) for the derived deficiencies. 
The normalized gas-phase abundances observed in the MB, derived from Equation~\ref{dg}, are given in Table~\ref{t5}
and plotted in Figure~\ref{fig6} for the total column density derived from the apparent optical depth
method. Table~\ref{t6} and Figure~\ref{fig7} show the relative abundance patterns for the two 
components at 179 and 198 \km. 

Previous ISM studies using high spectral resolution and high S/N ratio UV data
have shown a general progression of increasingly 
severe depletion from warm halo clouds, to warm disk clouds, to colder
disk clouds. Therefore, we compare the depletion in the MB to the depletion 
patterns observed in those three representative Galactic environments.
The Galactic results from previous UV studies 
\citep{jenkins87,savage96,fitzpatrick,welty99b} are summarized in 
Table~\ref{t5}. Note that the halo values for C, N, O, and Al from \citet{welty99b}
are estimated values since no actual measurements yet exist.
Figure~\ref{fig6} shows the comparison between 
the different Galactic environments and the results for the MB.
However, in order to compare the depletion 
pattern of the MB with those in the Galaxy, it may be necessary to correct the former for any 
differences in the underlying (undepleted) total elemental abundance for the MB and Galaxy. 
For the LMC and SMC abundances we have adopted the results of Russell and Dopita (1992,
and references therein, but see also Garnett 1999, Korn and Wolf 1999), 
except for Al (see discussion in Welty et al. 1997, 1999 and references therein),
and LMC Si abundance which is from \citet{korn}.
The last two columns in Table~\ref{t5} summarize these abundances, where the errors on
$[X/{\rm S}]_{\rm SMC/LMC}$ are typically $\pm 0.2$ dex.

In Figures~\ref{fig6} and \ref{fig7}, we indicate the corrections of the MB depletions assuming 
$(X/{\rm S})_{\rm MB} = (X/{\rm S})_{\rm SMC}$, as most of the MB gas may
arise from the SMC (Gardiner and Noguchi 1996, Rolleston et~al. 1999 
-- but see discussion in Section~\ref{discus}).
The MB depletions of Si, Fe, and Ni in the MB are very similar to the Galactic halo pattern, both
with or without these corrections (see Figure~\ref{fig6}). 
The component analysis in Figure~\ref{fig7} leads to the same conclusion for Si and Fe
for the individual components. The lower limits of \ion{C}{2}, \ion{N}{1}, and \ion{O}{1}
also are consistent with the Galactic depletion pattern. However, when the cloudlet sub-structure
is considered the non-saturated broader component of \ion{N}{1} appears
to be underabundant by $-1.3$ dex (or $-0.7$ dex when the SMC corrections are
applied). The \ion{Al}{2} lower limit is consistent with
the (estimated) Galactic halo depletion value.

\subsection{Temperature and density of the gas}
In principle, the component $b$-values can be used to determine the
kinetic temperature of the gas.  For the
sharper component of \ion{N}{1}, \ion{O}{1}, \ion{Si}{2} and \ion{Fe}{2}, 
the inferred temperature
has an uncertainty as large as the value itself. However, for
the narrower component of \ion{N}{1}, we can place an upper 
limit on the temperature, $T < 1.7 \times 10^4$ K. The detections of 
\ion{O}{1} and \ion{N}{1} suggest $T < 10^4$ K, since at higher temperatures 
most of the O and N would be ionized due to collisions \citep{sembach95}.

If collisions with electrons
are the principal source of excitation of \ion{C}{2*} in the MB, then 
 the familiar equation between \ion{C}{2} and \ion{C}{2*} \citep{spitzer}
can be written as
\begin{equation}\label{eq3}
n_e = \frac{1}{5.46} T^{0.5} \frac{n({\rm C}^{+*})}{n({\rm C}^{+})}\;\;\; {\rm cm}^{-3}\,.
\end{equation}
This relation provides an estimate of the electron density when the space 
densities 
are replaced by column densities, and therefore we make the standard 
supposition 
that the two excited states are spatially coincident along the sight line. 
\ion{C}{2}
is strongly saturated, so we use \ion{S}{2} as a proxy for \ion{C}{2} after 
scaling 
by the appropriate relative cosmic abundance and relative depletion of 
$-0.3$ dex \citep{spitzer},
so that $N({\rm C}^{+}) \sim   5 - 10 \times N({\rm S}^{+})$ (the factor 10 
is appropriate for
the Galaxy -- e.g. Spitzer and Fitzpatrick 1993, while the factor
of 5 accounts for the metallicity of the SMC -- Russell and Dopita 1992).
These ions are also the dominant ionization stages in both the neutral and ionized gas, 
and assuming that 
they are spatially coincident along the sight line,  
the electron density is:
$n_e \sim 5 $-$ 10 \times 10^{-4} T^{0.5} $ ${\rm cm}^{-3}$. For
$T < 1.7 \times 10^4$ K,  $n_e < 0.05-0.10$ ${\rm cm}^{-3}$. This density is 
derived from the total column densities of the ions, which are dominated by 
the strong, sharp component at 198 \km. The cloud at 179 \km is expected 
to be a low density gas as it is partially ionized (See Section~\ref{depdis}).

If ionization equilibrium applies ($\Gamma N(X^0)=\alpha n_e N(X^+)$),
the ratio of photoionization rates to recombination rates, $\Gamma/\alpha$,
can be estimated using the tentative measurement of \ion{C}{1}, the estimate
of \ion{C}{2} and the upper limit on $n_e$: $\Gamma/\alpha < 24 - 48$. For 
comparison, in the Galaxy $\Gamma/\alpha \approx 25$ ($T=100 \,{\rm K}$), 
$\Gamma/\alpha \approx 48$  ($T=300 \,{\rm K}$),  
$\Gamma/\alpha \approx 100$  ($T=1000 \,{\rm K}$) \citep[for the WJ1 radiation field]{pequignot}
which indicates that either the MB radiation field  is
lower than the typical Galactic field or that the temperature of the gas is low.

\subsection{Highly ionized species}
The detection of \ion{Si}{4} and \ion{C}{4} lines in the MB places
constraints on the temperature of the highly ionized species. 
In the Galactic halo, the high ionization absorption average 
ratios are: $N({\rm C^{+3}})/N({\rm Si^{+3}}) = 4.6 \pm 2.4$
and $N({\rm Si^{+3}})/N({\rm N^{+4}}) = 1.2 \pm 0.6 $ \citep{savage97}.
There is also hot gas in the SMC and LMC detected via 
\ion{Si}{4} and \ion{C}{4} absorption lines \citep{fitzpatrick85,deboer80}. 
The recent study of LMC coronal gas by \citet{wakker98} seems to suggest that
the processes for producing \ion{C}{4} may be similar in both galaxies. 

The widths of the MB \ion{Si}{4} and \ion{C}{4} lines imply a temperature of 
 $<1.7 \times 10^5$ K. As discussed previously, the process of deblending
the \ion{Si}{4} and \ion{C}{4} lines introduces some relatively large 
uncertainties in their derived column densities. Therefore, instead of 
considering the total column density results, we plot the ratio 
$N_a({\rm C^{+3}})/N_a({\rm Si^{+3}})$ versus the LSR velocity in 
Figure~\ref{fig8}.
The observed scatter and differences between the two \ion{Si}{4} 
$\lambda\lambda$1393, 1402 lines in this figure reflect primarily the uncertainties 
introduced by
deblending for the stellar lines. However, we find the average ratio 
$N_a({\rm C^{+3}})/N_a({\rm Si^{+3}}) \sim 1.9 \pm 0.9$ to be consistent with 
values observed along Galactic and extragalactic sight lines. \ion{N}{5}
is not detected in the DI\,1388 spectra (only the stellar lines are 
detected at 1238.82 and 1242.80 \AA), but the measured $3 \sigma$ upper limit 
of $\log N({\rm N^{+4}}) < 12.8$ dex (assuming a 
similar width to \ion{Si}{4} and \ion{C}{4} and that the lines are on
the linear part of the curve of growth) implies 
$N({\rm Si^{+3}})/N({\rm N^{+4}}) > 1.2$.  This value is compatible with
values observed in
the lower part of the Galactic halo and indicates that the production 
of the highly ionized species \citep{sembach92,sembach97,savage97}  
may be similar in both the Galactic halo and the MB. 

\section{Discussion}\label{discus}
\subsection{Interpretation of the depletion patterns}\label{depdis}
In the Galactic ISM, the underabundances of elements
along various sight line are usually attributed to the depletion of
these elements into dust. The nucleosynthetic history of the 
gas can also play a role, especially in a low metallicity medium such
as the MB. Russell and Dopita (1992; see also Welty et al. 1997, 1999 and references 
therein) showed that the relative abundances of the 
$\alpha$-elements (Ne, Mg, Si, S, Ar, Ca) and Fe-peak elements (Cr, Fe, Ni, Zn)
for the SMC and LMC \ion{H}{2} regions and young stars are similar to
those found in analogous Galactic objects. 
\citet{welty97,welty99}  presented detailed studies of two Magellanic 
Clouds sightlines (Sk 108 in the SMC, SN 1987A in the LMC), 
for which the gas-phase abundance patterns resemble those found either
for warm Galactic disk clouds or for clouds in the Galactic halo.  
They concluded that the similar gas-phase abundance patterns in the three
galaxies imply similar depletion patterns, despite global differences in 
their metal and dust content.

A detailed comparison of the relative stellar abundances in the MB with those in 
these three galaxies 
is not yet possible, as the derived absolute stellar abundances are poorly 
determined and are available only for C, N, O, Mg, and Si \citep{rol99}. 
However, 
differential abundance analyses of the young B-type stars in the MB show 
that, for those elements, the MB is deficient by $-1.1 \pm 0.2$ dex relative 
to the Galaxy.  The scatter
around this value is fairly small. This result has been confirmed recently
in a study of a supergiant star (Rolleston, priv. comm.) and in particular shows that 
the nitrogen abundance exhibits a similar deficiency
(see Section~\ref{nitro}). 
Our results show that the abundance pattern in the MB gas toward DI\,1388 
resembles that found in Galactic halo clouds and in the SMC gas toward Sk\,108.
Therefore, it seems that the dominant factor 
describing the MB 
gas-phase abundance pattern is depletion of the elements 
into dust rather than nucleosynthetic history.  Clearly, additional
studies of other MB sight lines would help to test this hypothesis.

The properties of the two MB ISM clouds toward DI\,1388 
can therefore be summarized as follows:

- {\underline {The 198 \km cloud}}: The gas-phase abundance pattern shown
in Figure~\ref{fig7} follows the 
Galactic halo pattern. The presence of strong \ion{O}{1}
and \ion{N}{1} suggests that much of the gas is neutral. There is no
detectable \ion{S}{3}, but
\ion{Si}{3} is detected, suggesting that some ionized gas is also present. This
ionized gas may be associated with the more highly ionized gas associated with
\ion{Si}{4} and \ion{C}{4}. This cloud consists of a low electron density, 
mainly neutral gas that might have a hotter, ionized boundary.

- {\underline {The 179 \km cloud}}: The gas-phase abundances of  \ion{Si}{2}
and \ion{Fe}{2} shown in Figure~\ref{fig7} indicate that this cloud also
follows a Galactic halo depletion pattern. However, \ion{N}{1} is remarkably 
deficient with respect to \ion{S}{2}, which probably indicates that the 
gas is partially ionized. This would also be consistent with the smaller 
deficiency of \ion{Si}{2} and \ion{Fe}{2} compared to \ion{N}{1}.
Similarly, \ion{O}{1} is only mildly saturated, which 
again may indicate that the gas is partially ionized. \ion{Si}{3} is present 
and saturated; but again indicating
that a certain amount of gas is partially ionized.  \ion{S}{3} is not detected,
but this is not surprising, as this feature is expected to be weak and below the detection
limit of our spectra.	

\subsection{Absolute metallicity of the Bridge gas}
Using the HIPASS 21 cm data with a spectral resolution of 1 \km and a
spatial resolution of $15.5\arcmin$, Putman (priv. comm.) found  
an \ion{H}{1} emission column density
$\log N({\rm H}^0) \simeq 20.30$ dex in the direction of DI\,1388. 
For comparison, the hydrogen column density obtained using 
\ion{S}{2} and scaling by the cosmic reference $[{\rm S}/{\rm H}] =-4.73$ dex 
is $\log N({\rm H}) \simeq 20.07$ dex after accounting for the 
metallicity of $-1.1$ dex.  This is in rough agreement with the 
column density derived from the 21 cm data. Differences could be due to the large beam of the 
\ion{H}{1} emission data and to the position (in depth along the sight line) 
of DI\,1388 in the MB.
However, our data generally support a metallicity of $\sim-1.1$ dex compared
to the Galaxy metallicity, confirming the results from the B-type 
stars study by \citet{rol99}. This implies that the metallicity
in the MB does not reflect the SMC metallicity, which is at odds with the
tidal model origin of the MB \citep{gardiner}. \citet{rol99} proposed 
that the MB gas was formed from a mixture of SMC gas and 
an unenriched component.

\subsection{The nitrogen ``problem"}\label{nitro}
The absolute nitrogen abundance in the SMC is still
very uncertain as \ion{H}{2} regions and stellar 
analyses yield substantially different results 
(about 0.5 dex, e.g. Garnett 1999, but see Dufton et~al. 1990 
for a counter-example). The reasons for this behaviour are still 
not well understood, mainly because the production mechanisms for N are still 
poorly constrained (see e.g., discussion in Russell and Dopita 1992, and references
therein). However, in principle it should be possible to 
supplement these analyses with interstellar data for the SMC 
and the MB since nitrogen is not readily incorporated into dust grains
\citep{meyer}.

Our data tentatively suggest that nitrogen is deficient in the MB gas
relative to other elements. For the component at 179 \km,
$[{\rm N^0}/{\rm S^+}] = -1.32 $ dex.  This deficiency is unlikely to be
due to depletion of N into grains. \ion{N}{1} compared to \ion{O}{1}
is also largely deficient ($<-0.5$ dex), which could be due in part
to \ion{N}{1} being preferentially ionized relative to O and H due 
its larger photoionization cross section. However, \citet{sofia98}
indicate that \ion{N}{1} is generally a good substitute for \ion{H}{1}
due to the very fast charge transfer of \ion{N}{2} with \ion{H}{1}
which effectively keeps the ionization fractions of N and  H essentially
coupled.  For the component at 198 \km, the saturated oxygen and 
nitrogen lines do not clarify the situation.
The amount of N that has been ionized
could be checked with  {\em Far Ultraviolet  Spectroscopic 
Explorer} measurements of \ion{N}{2} and \ion{N}{3} absorptions.
A far-UV study would also help to constrain the neutral nitrogen abundance,
as there are numerous \ion{N}{1} lines in the {\em FUSE} bandpass that are 
weaker than the strong 1200 \AA\ lines observed in our STIS spectrum of 
DI\,1388.

\section{Summary and concluding remarks}
We have presented {\em HST}\,  STIS E140M UV and AAT optical spectra of the 
young star DI\,1388 located in the Magellanic Bridge. The \'echelle spectra
show interstellar absorption from \ion{C}{1}, \ion{C}{2}, \ion{C}{2*}, 
\ion{C}{4}, 
\ion{N}{1}, \ion{O}{1}, \ion{Al}{2}, \ion{Si}{2}, \ion{Si}{3}, \ion{Si}{4}, 
\ion{S}{2}, \ion{Ca}{2},
\ion{Fe}{2}, and \ion{Ni}{2}. The relative gas-phase abundance of the MB
ISM gas toward DI 1388 resembles that of gas in the Galactic halo. Since there is 
independent information
on the total abundances from stars and/or nebulae of the Galaxy, SMC, LMC,
and MB, the observed
abundance pattern in the MB ISM is attributed to varying degrees of depletion 
onto dust similar to that in halo clouds and SMC toward Sk 108.

Fits to the absorption profiles reveal two MB interstellar components 
at 179 and 198 \km. 
These two clouds along the line of sight have different properties: 
(1) The cloud at 198 \km has a temperature $T< 10^4$ K, low electron density 
($n_e < 0.05$-0.1 ${\rm cm}^{-3}$), and is mainly neutral with a possible collisionally 
ionized boundary detected in \ion{Si}{3}, \ion{Si}{4}, and \ion{C}{4}. (2)
The cloud at 179 \km is low density gas, warmer, and partially ionized.

Higher resolution data might allow a more detailed picture of the interstellar 
structure, which bears on
the possibility that these clouds trigger 
star formation via cloud-cloud collisions \citep{dyson}. As discussed above, 
star formation is continuing in the MB, as there are young hot stars with 
evolutionary
lifetimes ($\la 20$ Myr) much younger than the age of the MB ($\sim 200$ Myr). 
In all known environments, star formation occurs within molecular clouds or 
cloud complexes, but there is presently no direct evidence of star-forming gas
clouds in the MB. \citet{kob} have detected cold \ion{H}{1} clouds in the MB, 
which suggests the presence of atomic or molecular condensations that could 
harbor
star formation. However, no CO emission has been yet detected
\citep{smoker},  and therefore cloud-cloud collisions could be the 
dominant star formation trigger. 

Finally, we note that this work is relevant to the study of QSO 
absorption-line systems (QSOALS).
\citet{welty97,welty99} propose that the known depletions patterns in the SMC, LMC, and Galaxy 
may be used to deduce the total elemental abundances in the QSOALS from the 
observed gas-phase abundances. Indeed, the MB interstellar gas is 
characterized by modest depletion and metallicity.  Moreover, the results of
this study and those of \citet{welty97,welty99}  suggest 
that the observed depletion patterns are independent of the metallicity.
These patterns could therefore be applied to the QSOALS. 
Detailed studies of the interstellar medium along other sight lines in the MB, SMC, 
and LMC should help to know whether these results can be generalized to a wider range of environments.

\acknowledgments
We are grateful to Mary Putman for providing the HIPASS data toward the 
general direction of DI\,1388 and for useful discussions.
We appreciate insightful comments from Chris Howk.
NL held a postgraduate studentship from the European Social Fund and
Northern Ireland Development for Research through most time of this work. 
WRJR acknowledges financial support
from the UK Particle and Astronomy Research Council.
This research has made use of the NASA
Astrophysics Data System Abstract Service (\url{http://adswww.harvard.edu/})
and the CDS database (\url{http://cdsweb.u-strasbg.fr/}).
We thank Dan Welty for his careful reading of the manuscript and 
insightful comments.

\newpage
\begin{table*}[!t]
\begin{center}
\caption{Atmospheric and other observational parameters of DI\,1388}
\label{t0}
\begin{tabular}{lccccccccc}
\tableline\tableline
Star     &  $\alpha$ &  $\delta $ &  $V$   & $B-V$ & $T_{\rm eff}$ & $\log g$ & $v_{\rm t}$  & $v^\star_{\rm LSR}$ & $v\sin i$ \\
	& (J2000) & (J2000) & & & kK & dex & \km & \km & \km 
\\
\tableline
DI\,1388  	 &  02:57:12.94 &   -72:52:54.6  & 14.39 & $-0.26 $ & $32 \pm 1$ &  $4.0 \pm 0.2$  & $5\pm 5$ & $150 \pm 30$ & $180 \pm 30$ \\
\tableline
\end{tabular}
\tablecomments{Results are from \citet{hambly} and references therein.}
\end{center}
\end{table*}

\begin{deluxetable}{lcccccc}
\tablecolumns{6}
\tablewidth{0pc} 
\tablecaption{List of observed interstellar absorption lines in the DI\,1388 spectrum and their total equivalent widths and  
column densities \label{t3}} 
\tablehead{\colhead{Ions}    &   \colhead{$\lambda_{\rm lab}$\tablenotemark{a}} &   \colhead{$f$\tablenotemark{a}} &
\colhead{S/N} &\colhead{$W_{\rm MB}$} & \colhead{$\log N_{\rm MB}$} &  \colhead{Note}\\ 
\colhead{}  &\colhead{\AA} & \colhead{} &\colhead{} &\colhead{m\AA}& \colhead{${\rm cm}^{-2}$} & \colhead{}}
\startdata
\ion{C}{1}  	 &  1277.245 &   $9.67 \times 10^{-2}$    &	27		          	& $< 6.0$				  & $< 12.63 $  	      	       &      \\
 		 &  1328.833 &   $6.30 \times 10^{-2}$ \tablenotemark{b}   &	33           	& $< 5.2$				  & $< 12.72 $         			&       \\
  		 &  1560.309 &   $8.04 \times 10^{-2}$ &	22  		          	& $< 9.4$				  & $< 12.74 $  	    	       & 	\\
 		 &  1656.928 &   $1.40 \times 10^{-1}$    &	14   		          	& $< 15.4$				  & $< 12.65 $  	    	       & 	\\
\ion{C}{2} 	 &  1334.532 &   $1.28 \times 10^{-1}$    &	36   		          	& $232.7 \pm 10.5$			  & $> 14.87 $  	   	       &   1    \\
\ion{C}{2*} 	 &  1335.708 &   $1.15 \times 10^{-1}$    &	38   		          	& $9.9 \pm 2.3$ 			  & $ 12.65 \pm 0.06$	     	       &        \\
\ion{C}{4} 	 &  1550.770 &   $9.52 \times 10^{-2}$    &	22  		          	& $23.3 \pm 5.9$			  & $ 12.90 \pm 0.28$		       &   2  \\
\ion{N}{1} 	 &  1199.550 &   $1.33 \times 10^{-1}$    &	17   		          	& $53.1 \pm 3.9$			  & $ >14.03 $ 		 	       &      \\
  		 &  1200.223 &   $8.85 \times 10^{-2}$    &	29   		          	& $42.0 \pm 3.3$			  & $ > 13.82 $	      		       &      \\
  		 &  1200.710 &   $4.42 \times 10^{-2}$    &	21   		          	& $40.2 \pm 3.5$			  & $ > 14.01 $	      		       &      \\
\ion{O}{1}  	 &  1302.168 &   $4.89 \times 10^{-2}$    &	43   		          	&$168.9 \pm 10.1$			  & $ > 14.83 $ 	    	       & 1	\\
\ion{Mg}{2} 	 &  1239.925 &   $6.17 \times 10^{-4}$\tablenotemark{c}   &	19        	& $< 8.4$				  & $< 15.00 $  	     	       &       \\
  		 &  1240.395 &   $3.54 \times 10^{-4}$\tablenotemark{c}   &	19        	& $< 8.4$				  & $< 15.24 $  	     	       &       \\
\ion{Al}{2} 	 &  1670.787 &   $1.83 $	           &	14   		          	&$190 :$			 	  & $ > 12.82$		      	       & 3	\\
\ion{Si}{2} 	 &  1190.416 &   $2.50 \times 10^{-1}$    &	14   		          	&$140.1\pm 11.8$ 			  & $> 14.07 $  	      	       &      \\
 		 &  1193.290 &   $4.99 \times 10^{-1}$    &	20   		          	&$170.8 \pm 12.3$			  & $> 13.90 $  	      	       &      \\
		 &  1260.422 &   $1.01 $	           &	24   		          	&$220.4 \pm 18.6$			  & $ 14.23 : $			       &  1    \\
 		 &  1304.370 &   $8.60 \times 10^{-2}$\tablenotemark{d}   &	33 	  	&$113.2 \pm 11.8$			  & $ 14.20 \pm 0.08$	      	       &      \\
 		 &  1526.706 &   $1.10 \times 10^{-1}$\tablenotemark{d}   &	24  	  	&$155.8 \pm 14.0$			  & $ 14.17 \pm 0.11$	      	       &      \\
\ion{Si}{3}      &  1206.500 &   $1.67 $	           &	28   		          	&$217.3 \pm 13.8$			  & $> 13.45 $			       & 1,2     \\
\ion{Si}{4}      &  1393.755 &   $5.14 \times 10^{-1}$    &	32   		          	& $59.0 \pm 5.6$			  & $ 12.86 \pm 0.08$		       &  2    \\
 		 &  1402.770 &   $2.55 \times 10^{-1}$    &	27   		          	& $35.9 \pm 6.1$			  & $ 12.95 \pm 0.07$		       &  2   \\
\ion{P}{2}       &  1152.818 &   $2.36 \times 10^{-1}$    &	 5   		          	& $< 31.2$				  & $< 13.05 $  	      	       &      \\
\ion{S}{2}       &  1250.584 &   $5.45 \times 10^{-3}$    &	29   		          	&$ 13.1 \pm 1.4$			  & $ 14.20 \pm 0.14$	      	       &       \\
  		 &  1253.811 &   $1.09 \times 10^{-2}$    &	27   		          	& $25 \pm 3.1$			 	  & $ 14.28 \pm 0.13$	      	       &      \\
 		 &  1259.519 &   $1.62 \times 10^{-2}$    &	24   		          	& \nodata 				  &  \nodata 		       &  1    \\
\ion{S}{3}       &  1190.208 &   $2.22 \times 10^{-2}$    &	14   		          	&$ < 10.8$				  & $ < 13.59$         		       &      \\
\ion{Ca}{2} 	 &  3934.777 &   $6.35 \times 10^{-1}$    &	39   		          	& $50.3 \pm 2.0$			  & $11.75 \pm 0.09 $		       &      \\
\ion{Mn}{2}      &  1197.184 &   $1.57 \times 10^{-1}$    &	20   		          	& $< 11.8$				  & $< 12.78 $  	    	       & 	\\
\ion{Fe}{2} 	 &  1260.533 &   $2.50 \times 10^{-2}$    &	33   		          	&$ 36 :$				  & $ 14.09: $			       & 1     \\
 		 &  1608.451 &   $5.80 \times 10^{-2}$\tablenotemark{e}   &	20        	&$83.0 \pm 9.0$			 	  & $ 13.91 \pm 0.03$	     	       &       \\
 		 &  1611.200 &   $1.02 \times 10^{-3}$\tablenotemark{f}   &	19        	& $< 11.0$				  & $< 14.67 $  	      	       &      \\
\ion{Ni}{2} 	 &  1317.217 &   $7.80 \times 10^{-2}$\tablenotemark{g}    &	30  		& $< 5.8$				  & $< 12.68 $  	      	       &      \\
 		 &  1370.132 &   $7.70 \times 10^{-2}$\tablenotemark{g}    &	39           	& $5.2 \pm 4.7$ 			  & $ 12.66 \pm 0.20$		         &  4   \\
 		 &  1454.842 &   $2.98 \times 10^{-2}$\tablenotemark{g}    &	31            	&$5.5 \pm 5.2$  			  & $ 12.57 \pm 0.40$		         &  4   \\
\enddata
\tablecomments{Uncertainties are $1\sigma$ error (see text). Upper limits indicate that no feature is present and
are $3 \sigma$ estimates (see text for more details). Lower limits indicate that the absorption line is 
saturated. A colon indicates that the value is uncertain.
(1) Line blended with another interstellar absorption line or another velocity component.
(2) Line is blended with the stellar line.
(3) The MB component is only observed partly ($\sim 70-80\%$).
(4) Line very weak.}
\tablenotetext{a}{Rest frame vacuum wavelengths and oscillator strengths are from Morton (1991), unless otherwise stated.}
\tablenotetext{b}{From \citet{wiese}.}
\tablenotetext{c}{From \citet{sofia}.}
\tablenotetext{d}{From \citet{spitzer,dufton83,dufton92}.}
\tablenotetext{e}{From \citet{mullman}.}
\tablenotetext{f}{From \citet{cardelli}.}
\tablenotetext{g}{From \citet{fedchak,zsargo}.} 
\end{deluxetable}

\begin{deluxetable}{lccccc}
\tablecolumns{4}
\tablewidth{0pc} 
\tablecaption{Component fitting measurements at $\sim 179$ and $\sim 198$ \km \label{t4}} 
\tablehead{ \colhead{Ions}    &   \colhead{$\lambda_{\rm lab}$ (\AA)}  &  \multicolumn{2}{c}{$b$ (\km)} & \multicolumn{2}{c}{$\log N$ (${\rm cm}^{-2}$)} \\ 
\cline{3-4} \cline{4-6} \\ 
\colhead{}    &   \colhead{}  &  \colhead{179 \km} & \colhead{198 \km}&  \colhead{179 \km} & \colhead{198 \km}}
\startdata
\ion{C}{2}	 &  1334.532 &$12.0: $& $8.0:$  		   	& $ > 14.54 $& $ > 14.60$	    \\
\ion{N}{1}	 &  1200.223 &\nodata& $4.5 \pm 0.6$  		   	& \nodata & $ 14.00 \pm 0.12$	    \\
  		 &  1200.710 &$11 : $& $5.1 \pm 0.5$			& $ 13.05 \pm 0.19$& $ > 14.07 $	    \\
\ion{O}{1}  	 &  1302.168 &$10.7 \pm 0.3$& $6.7 \pm 0.3$     	& $ > 14.44 $& $ > 14.51 $		    \\
\ion{Al}{2}	 &  1670.787 &$11.7:$& $8.7:$     			& $ > 12.5: $	& $ > 12.6: $\\
\ion{Si}{2}	 &  1304.370 &$10.9 \pm 0.4$& $7.6 \pm 0.4$     	& $ 13.68 \pm 0.06$& $ 14.06 \pm 0.03 $\\
 		 &  1526.706 &$10.6 \pm 0.4$& $6.6 \pm 0.4$     	& $ 13.73 \pm 0.07$& $ 14.00 \pm 0.05 $	    \\
\ion{S}{2}       &  1250.584 &$7.1 \pm 1.2$& $5.1 \pm 1.2$     	& $ 13.67 \pm 0.15$& $ 14.11 \pm 0.08 $  \\
\ion{Ca}{2} 	 &  3933.663 &$12.7 \pm 0.7$& $7.8 \pm 0.6$     	& $ 10.93 \pm 0.17$& $ 11.73 \pm 0.05 $	      \\
\ion{Fe}{2}	 &  1608.451 &$10.0 \pm 0.5$& $6.0 \pm 0.5$     	& $ 13.38 \pm 0.05$& $ 13.77 \pm 0.04 $\\
\enddata
\end{deluxetable}

\begin{deluxetable}{lcccccccc}
\tablecolumns{9}
\tablewidth{0pc} 
\tablecaption{Summary of depletions of the MB toward  DI\,1388\label{t5}} 
\tablehead{ \colhead{Ions}    &   \colhead{}&   \colhead{} &   \colhead{MB}   &  \multicolumn{3}{c}{Galactic Depletions} 
& \colhead{SMC} & \colhead{LMC} \\ 
\cline{5-7} \\ 
\colhead{$X^i$}  & \colhead{$\log\left(\frac{X}{{\rm S}}\right)_{\rm c}^a $} &   \colhead{$\log (N(X^i))$$^b$} & 
\colhead{$D(X)$}    & \colhead{Cold$^c$}   & \colhead{Warm$^c$}    & \colhead{Halo$^d$}& \colhead{$[X/{\rm S}]^e$} & \colhead{$[X/{\rm S}]^e$}}
\startdata
\ion{C}{2}  	&  $+1.28$    &$> 14.87$ & $ > -0.65$  		        & $-0.4$& $-0.4$& $(-0.4)$ 		& $-0.14$& $+0.06$\\
\ion{N}{1}	&  $+0.70$    &$> 14.00$ & $> -0.94 $ 		        & $-0.1$& $-0.1$& $(-0.1)$		& $-0.66$& $-0.26$ \\
\ion{O}{1}	&  $+1.60$    &$> 14.83$ & $> -1.01$  		        & $-0.4$& $-0.4$& $(-0.4)$ 		& $-0.16$& $+0.05$\\
\ion{Mg}{2}	&  $+0.31$    &$< 15.00$ & $< +0.45$     		& $-1.2$& $-0.6$& $-0.3$ & 		  $+0.08$& $+0.44$\\
\ion{Al}{2}	&  $-0.79$    &$> 12.82$ & $> -0.63$     		& $-2.4$& $-1.1$& $(-0.6)$ 		& $+0.20$& $+0.19$\\
\ion{Si}{2}	&  $+0.28$    &$  14.19 \pm 0.14$ & $-0.33 \pm 0.16$    	& $-1.3$& $-0.4$& $-0.3$ 	& $+0.16$& $+0.02$\\
\ion{P}{2}	&  $-1.70$    &$< 13.05$ & $<+ 0.51$    		& $-0.5$& $-0.2$& $(-0.1)$ 		& \nodata& \nodata\\
\ion{Mn}{2}	&  $-1.74$    &$< 12.78$ & $< +0.28$    		& $-1.5$& $-1.0$& $-0.7$ 		& $+0.18$& $+0.25$\\
\ion{Fe}{2}	&  $+0.24$    &$  13.91 \pm 0.05$ & $-0.57 \pm 0.19 $   	& $-2.2$& $-1.4$& $-0.6$ 	& $+0.01$& $+0.29$\\
\ion{Ni}{2}	&  $-1.02$    &$  12.52 \pm 0.20$ & $-0.70 \pm 0.25 $   	& $-2.2$& $-1.4$& $-0.6$ 	& $+0.28$& $+0.36$\\
\enddata
\tablecomments{$(a)$ Solar system meteoritic abundances from \citet{anders} except for C, N and O, which are
photospheric values from \citet{grevesse}; $\log({\rm S/H})_{\rm c} = -4.73$.$(b)$ Adopted \ion{S}{2}
column density, $\log(N({\rm S^+}))= 14.24 \pm 0.19$ dex. $(c)$ Updated 
from \citet{jenkins87}, see \citet{lauroesch,welty97,welty99b}. $(d)$ From \citet{savage96,fitzpatrick}; and references
therein. Values in parentheses are estimated, see \citet{welty97,welty99b}.
For \ion{Ni}{2}, the depletions were corrected to take into account the new oscillator strength scaling.
$(e)$ $[X/{\rm S}]= \log(X/{\rm S})_{\rm SMC/LMC} - \log(X/{\rm S})_c$ (\citet{russell} but adjusted for C, N and O
photospheric values, and Al abundances are from \citet{welty97,welty99}, while LMC Si abundance is from \citet{korn},
but see also discussion in \citet{garnett}).}
\end{deluxetable} 

\begin{deluxetable}{lcccccc}
\tablecolumns{7}
\tablewidth{0pc} 
\tablecaption{Observed depletions for the two MB clouds at 179 and 198 \km. \label{t6}} 
\tablehead{ \colhead{Ions}    &   \colhead{}&    \multicolumn{2}{c}{MB Depletions}   &  \multicolumn{3}{c}{Galactic Depletions$^a$} \\ 
\cline{5-7} \\ 
\colhead{$X^i$}  & \colhead{$\log\left(\frac{X}{{\rm S}}\right)_{\rm c} $}  & 
\colhead{$D_{179}(X)$} & \colhead{$D_{198}(X)$}    & \colhead{Cold}   & \colhead{Warm}    & \colhead{Halo}}
\startdata
\ion{C}{2}  	& $+1.28$   & $>-0.41$ & $>-0.79 $    & $-0.4$& $-0.4$& $(-0.4)$ \\
\ion{N}{1}	& $+0.70$   & $-1.32 \pm 0.23$ & $>-0.79 $    & $-0.1$& $-0.1$& $(-0.1)$ \\
\ion{O}{1}	& $+1.60$   & $> -0.83  $ & $>-1.20 $	 & $-0.4$& $-0.4$& $(-0.4)$ \\
\ion{Al}{2}	& $-0.79$   & $ >-0.43 :$ & $ >-0.74 :$    & $-2.4$& $-1.1$& $(-0.6)$ \\
\ion{Si}{2}	& $+0.28$   & $-0.25 \pm 0.16$ & $-0.36 \pm 0.09$    & $-1.3$& $-0.4$& $-0.3$  \\
\ion{Fe}{2}	& $+0.24$   & $-0.53 \pm 0.16$ & $-0.58 \pm 0.09$    & $-2.2$& $-1.4$& $-0.6$  \\
\enddata
\tablecomments{($a$) Values in parentheses are estimated. See footnote in Table~\ref{t5} and text for references.}
\end{deluxetable}

\clearpage
\begin{figure*}[!th]
\begin{center}
\includegraphics[width=16.5 truecm]{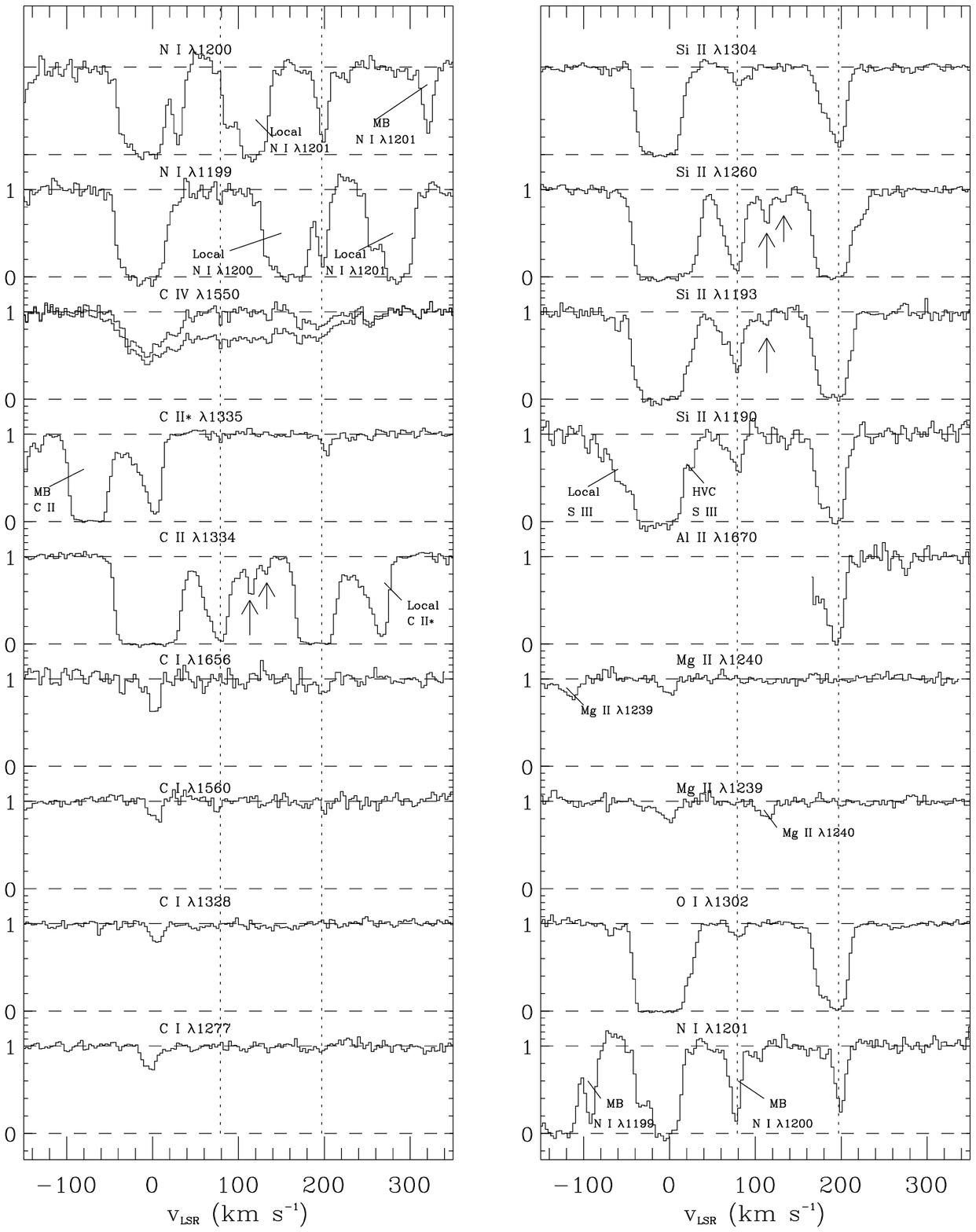}
\caption{Normalized profiles of absorption lines in the {\em HST}\, STIS spectrum of DI\,1388 plotted
against LSR velocity. Absorption from the Magellanic Bridge is detected at $\sim 200$ \km, while 
the HVC discussed by \citet{lehner1} is at $\sim 80$ \km. The arrows indicate the detection of two 
other weak HVCs at 113 and 130 \km. For \ion{C}{4} and \ion{Si}{4}, we have 
indicated the observed line profile (lower profile) and the interstellar feature deblended from the stellar 
line spectrum (see text for more detailed information).} 
\label{fig2}
\end{center}
\end{figure*}

\begin{figure*}[!th]
\begin{center}
\includegraphics[width=16.5 truecm]{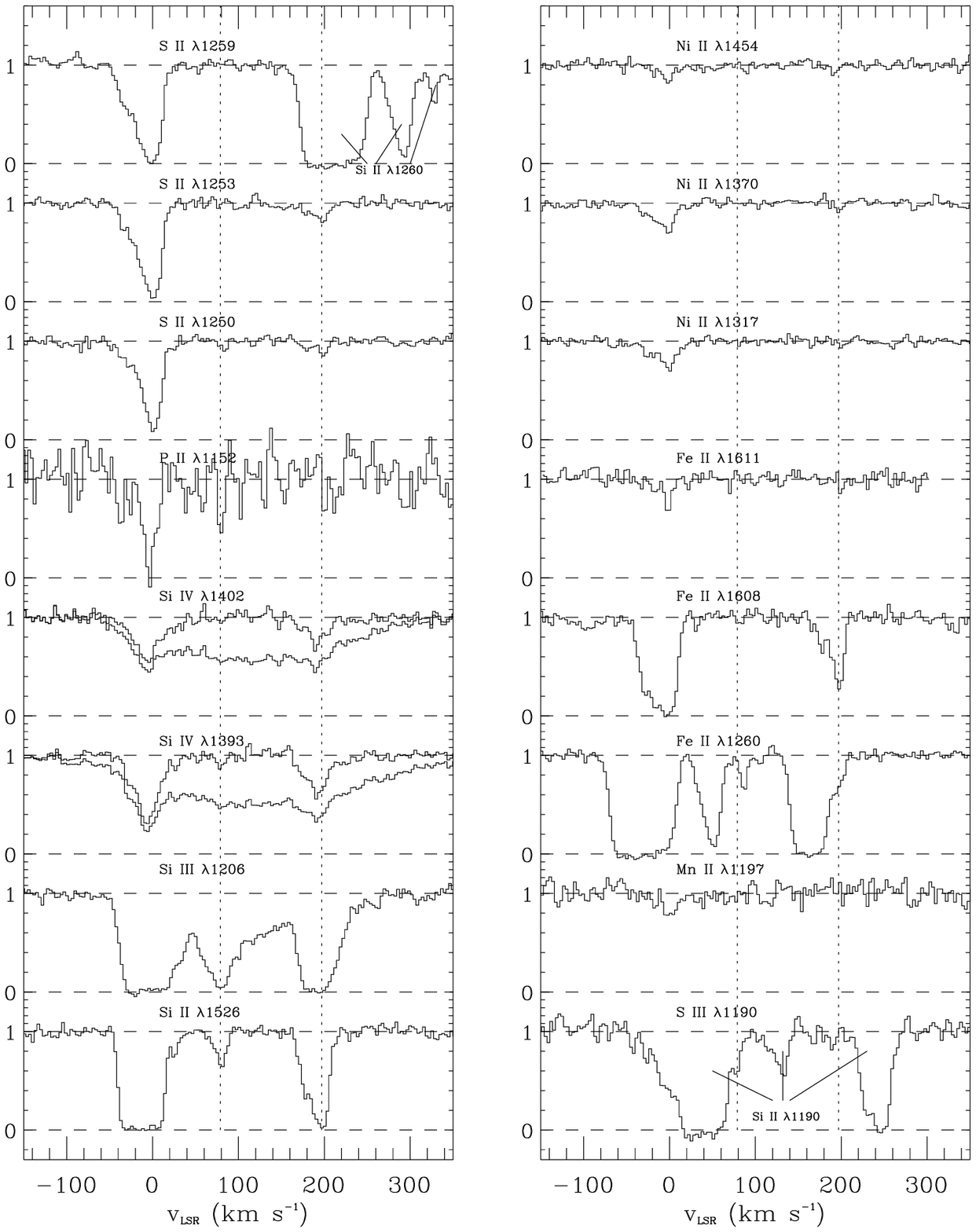}
\figurenum{\ref{fig2}}
\caption{continued.}
\end{center}
\end{figure*}

\begin{figure*}[!t]
\begin{center}
\includegraphics[width=16 truecm]{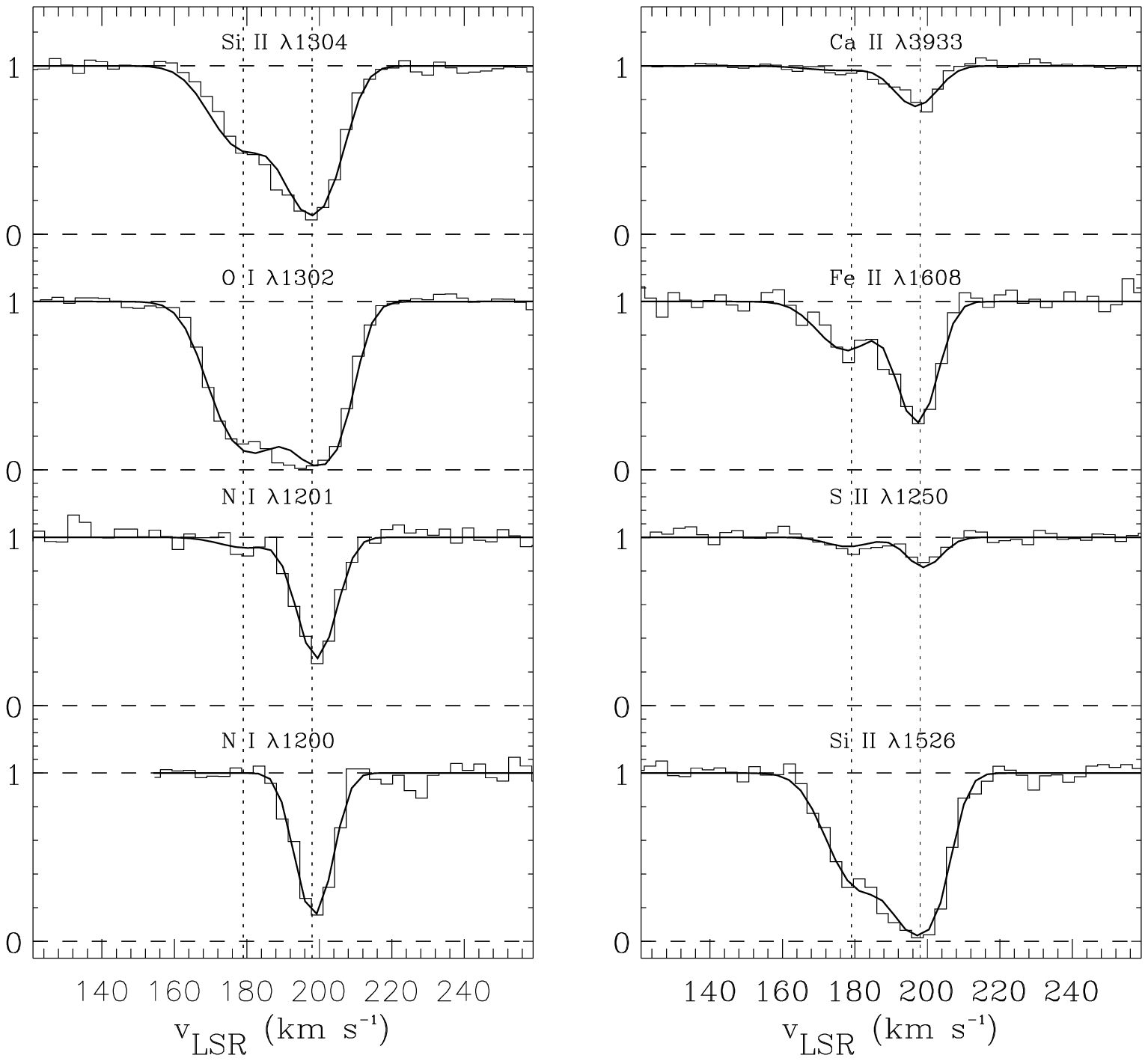}
\caption{Histograms are the observed spectra. Smooth lines are the fits with two Voigt 
components (except for \ion{N}{1} $\lambda$1200 where one component is sufficient). Dotted lines show the 
component centroids at 179 and 198 \km.  The fit for \ion{O}{1}
is relatively poor in comparison with the others, indicating that the sharp component 
has some unresolved saturated structures or perhaps additional weak components.\label{fig3}} 
\end{center}
\end{figure*}

\begin{figure*}[!t]
\begin{center}
\includegraphics[width=15 truecm]{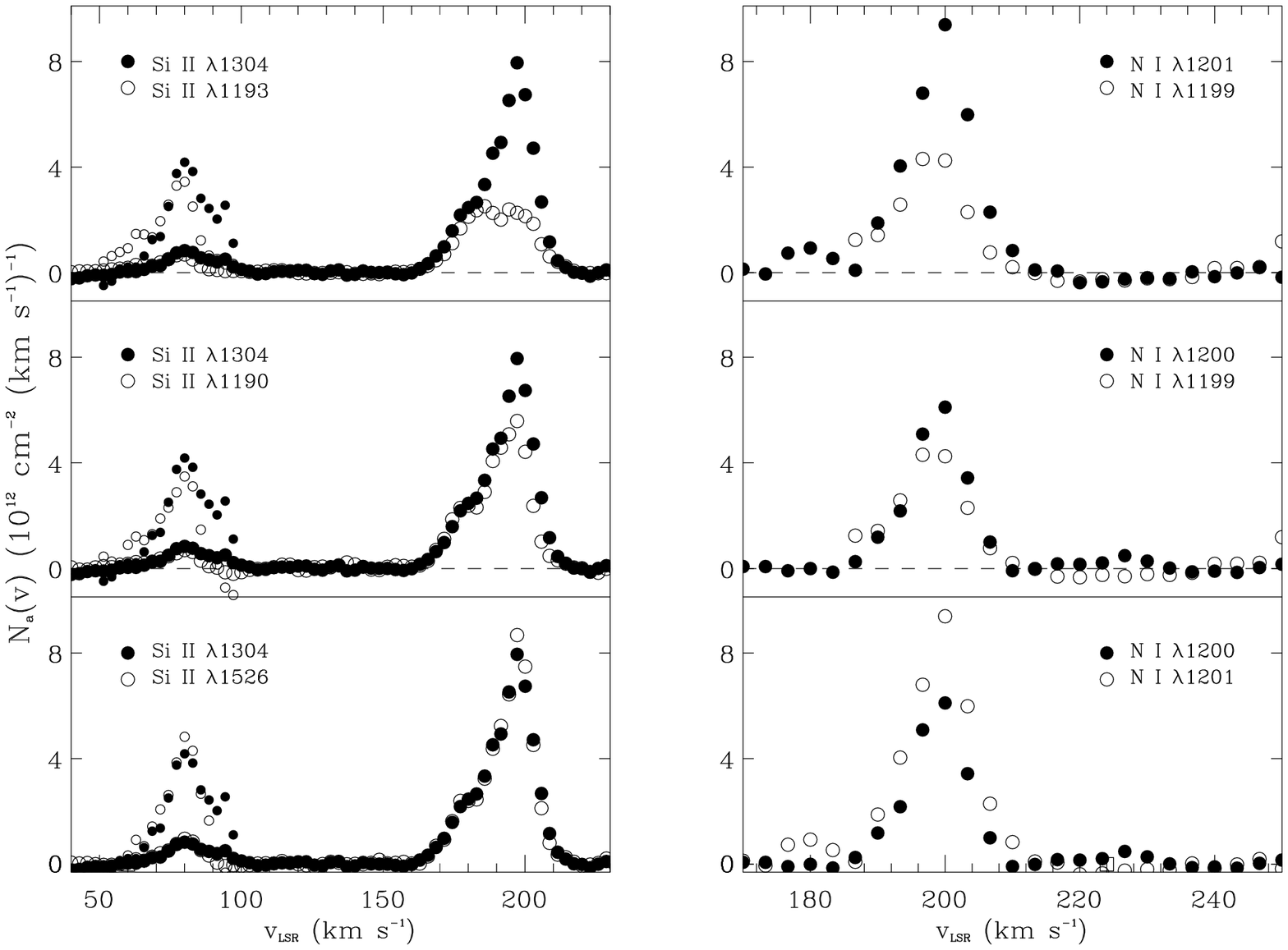}
\caption{Profiles of apparent column density of \ion{Si}{2} and \ion{N}{1} 
versus $v_{\rm LSR}$ at the HVC (80 \km) and MB (200 \km) absorption toward DI\,1388.  The smaller circles
at the velocities of the HVC absorption show the same profiles expanded by a factor of
5 in the vertical scale. The MB \ion{N}{1} $\lambda$1199 is blended with the local \ion{N}{1} $\lambda$1200:
therefore for $\lambda$1199 transition, we only plot the apparent column density when 
$v_{\rm LSR} > 180$ \km.} 
\label{fig4}
\end{center}
\end{figure*}

\begin{figure*}[!t]
\begin{center}
\includegraphics[width=8 truecm]{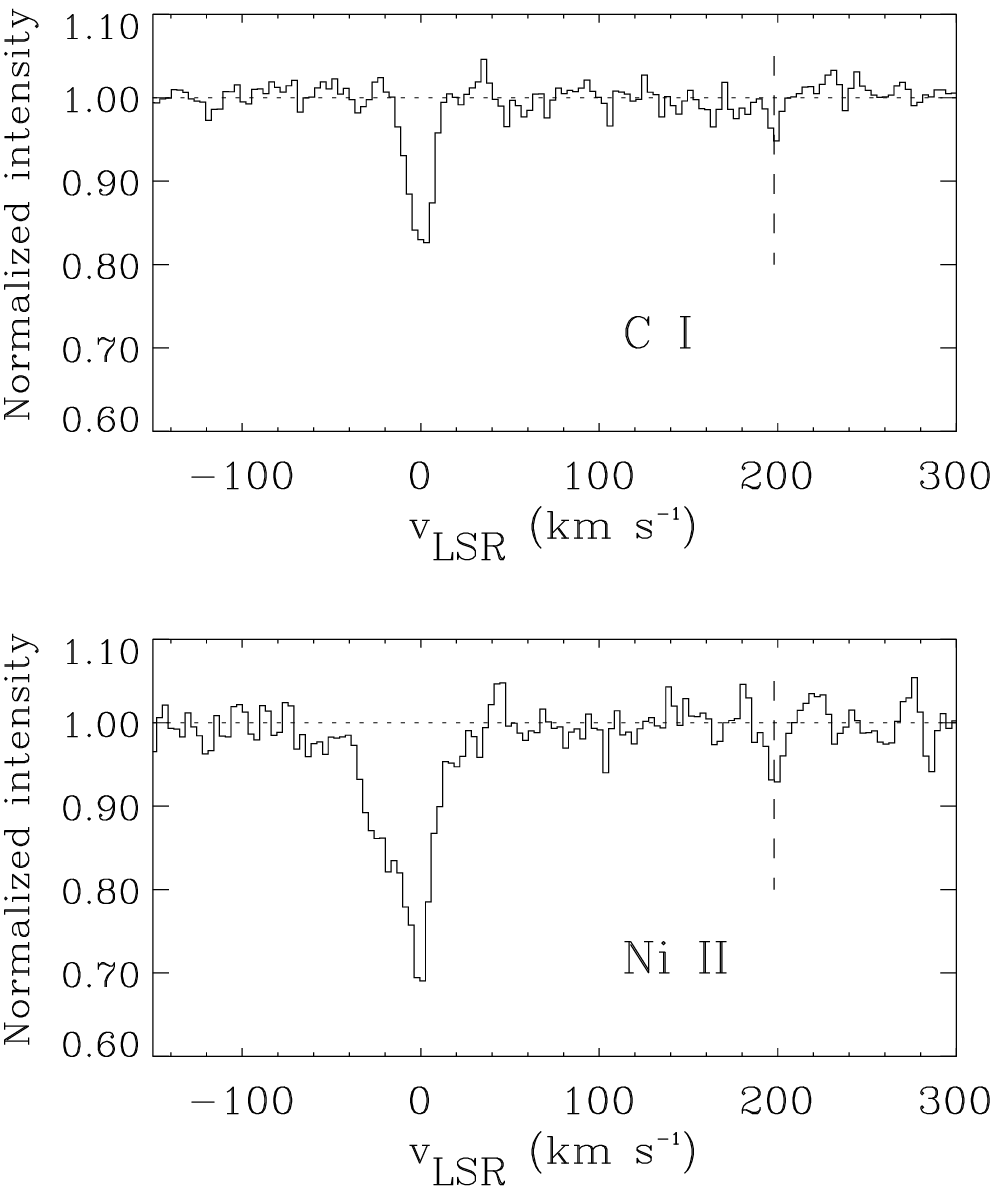}
\caption{Histograms of the co-added spectra for \ion{C}{1} and  \ion{Ni}{2}.} 
\label{fig9}
\end{center}
\end{figure*}

\begin{figure*}[!t]
\begin{center}
\includegraphics[width=12 truecm]{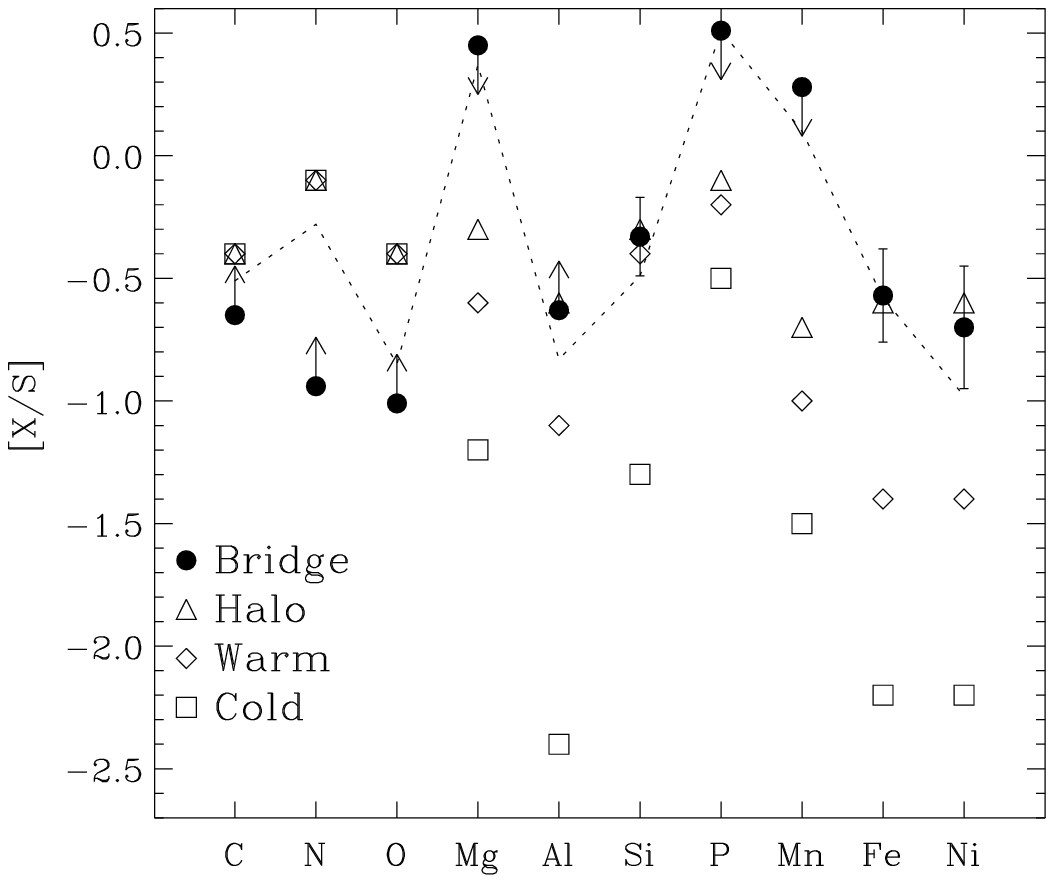}
\caption{Relative gas-phase abundances for the MB (DI\,1388 sight line), with respect to
sulfur, compared to relative abundances found in Galactic cold, warm and halo clouds.
Error bars are $1\sigma$. The upward (downward) arrows indicate lower (upper) limits. The Dotted line
represents the MB depletions corrected from the  underlying (undepleted) total elemental relative 
abundance patterns of the SMC with respect to the Galaxy. \label{fig6}} 
\end{center}
\end{figure*}

\begin{figure*}[!t]
\begin{center}
\includegraphics[width=12 truecm]{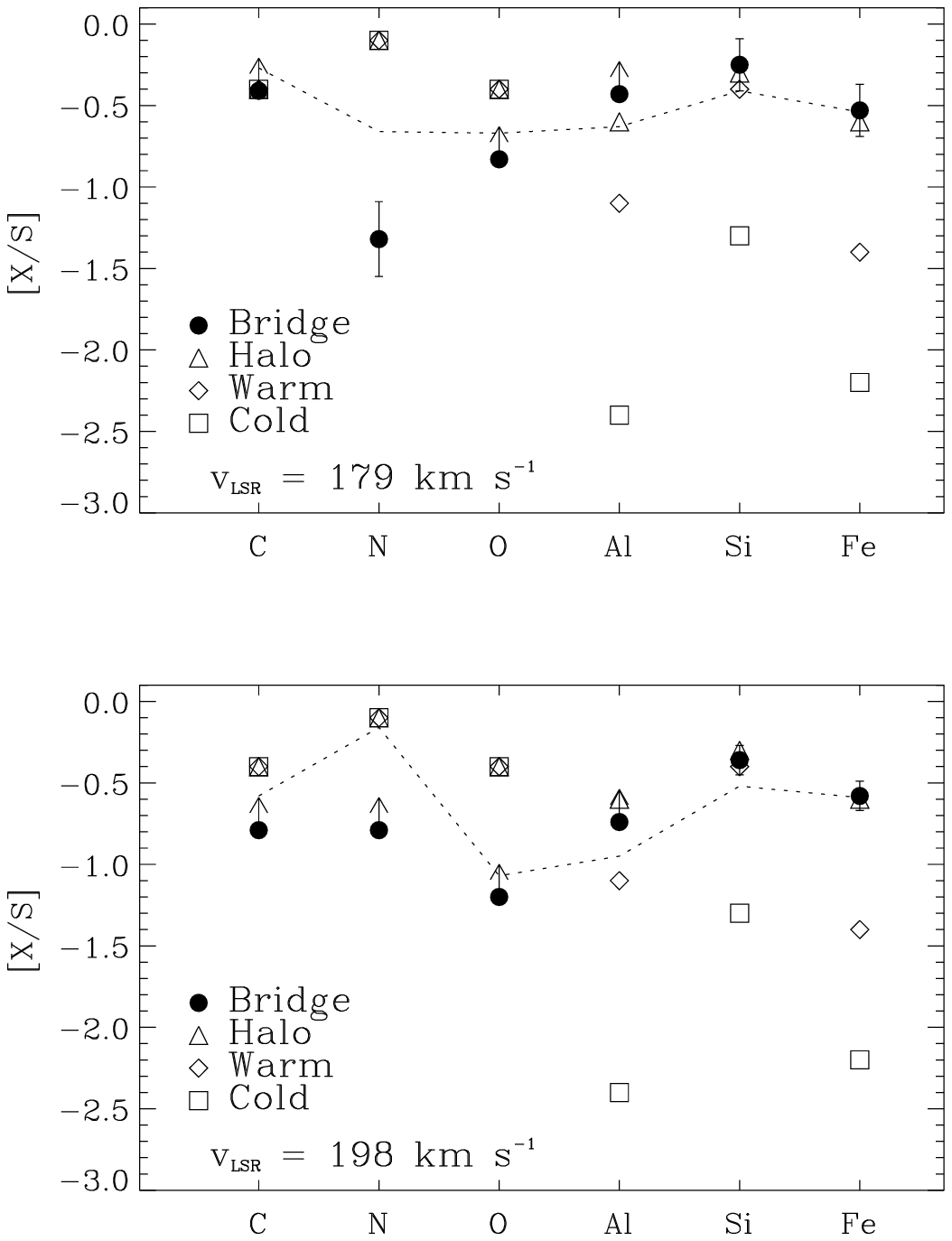}
\caption{Same as Figure~\ref{fig6}, but for the two clouds at 179 and 198 \km.  \label{fig7}} 
\end{center}
\end{figure*}

\begin{figure*}[!t]
\begin{center}
\includegraphics[width=12 truecm]{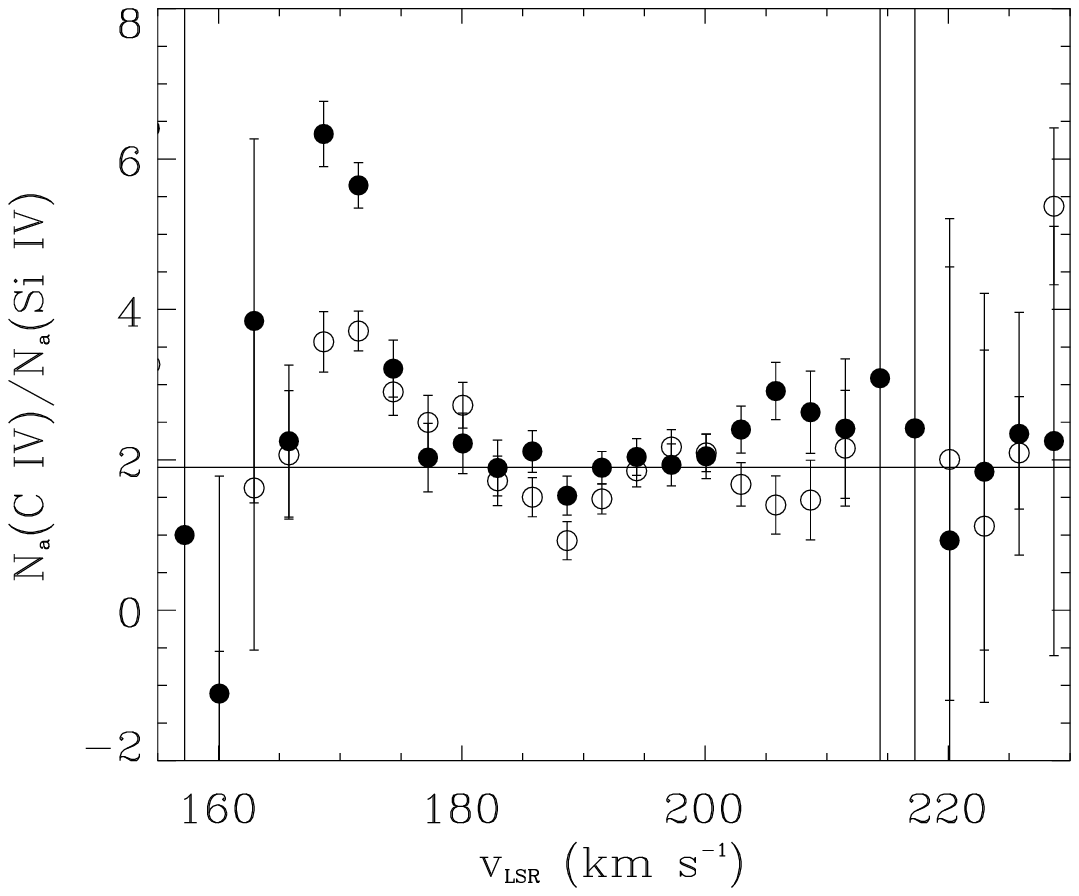}
\caption{$N_a({\rm C^{+3}})/N_a({\rm Si^{+3}})$ line ratio versus LSR velocity
in the MB component.  Open circles correspond
to the comparison of \ion{C}{4} $\lambda$1550 with \ion{Si}{4} $\lambda$1393, and filled circles correspond
to the comparison pf \ion{C}{4} $\lambda$1550 with \ion{Si}{4} $\lambda$1402. The solid line is the mean value of this ratio
in the velocity range of 180 and 200 \km. \label{fig8}} 
\end{center}
\end{figure*}

\end{document}